\documentclass[prl,twocolumn,notitlepage,showpacs, superscriptaddress,aps,amsmath,amssymb,UFT8]{revtex4-1}
\usepackage[usenames,dvipsnames]{color} %for color
\usepackage{graphicx}% Include figure files
\usepackage{CJK}
\definecolor{LinkColor}{rgb}{0,0,.5}
\usepackage[colorlinks=true,linkcolor=BrickRed,citecolor=LinkColor,urlcolor=LinkColor]{hyperref}
\usepackage{multirow}
\usepackage{comment}

\newcommand{\carb}{$^{13}$C }
\newcommand{\Nit}{$^{14}$N }
\newcommand{\ket}[1]{\left\vert{#1}\right\rangle}

\newcommand{\braket}[2]{\langle{#1}\vert{#2}\rangle}

\newcommand{\NVm}{NV$^-$}
\newcommand{\NVz}{NV$^0$}

\begin{document}
\begin{CJK*}{UTF8}{} 

\title{Repetitive Readout Enhanced by Machine Learning}% Force line breaks with \\%
\author{Genyue Liu \CJKfamily{gbsn}(刘亘越)}
\thanks{These authors contributed equally to this work.}
\affiliation{
Research Laboratory of Electronics, Massachusetts Institute of Technology, Cambridge, Massachusetts 02139, USA
}

\author{Mo Chen \CJKfamily{gbsn}(陈墨) }
\thanks{These authors contributed equally to this work.}
\affiliation{
Research Laboratory of Electronics, Massachusetts Institute of Technology, Cambridge, Massachusetts 02139, USA
}
\affiliation{
Department of Mechanical Engineering, Massachusetts Institute of Technology, Cambridge, Massachusetts 02139, USA
}

\author{Yi-Xiang Liu \CJKfamily{gbsn}(刘仪襄)}
\affiliation{
Research Laboratory of Electronics, Massachusetts Institute of Technology, Cambridge, Massachusetts 02139, USA
}
\affiliation{
Department of Nuclear Science and Engineering, Massachusetts Institute of Technology, Cambridge, Massachusetts 02139, USA
}

\author{David Layden}
\affiliation{
Research Laboratory of Electronics, Massachusetts Institute of Technology, Cambridge, Massachusetts 02139, USA
}
\affiliation{
Department of Nuclear Science and Engineering, Massachusetts Institute of Technology, Cambridge, Massachusetts 02139, USA
}

\author{Paola Cappellaro }
\thanks{pcappell@mit.edu}
\affiliation{
Research Laboratory of Electronics, Massachusetts Institute of Technology, Cambridge, Massachusetts 02139, USA
}
\affiliation{
Department of Nuclear Science and Engineering, Massachusetts Institute of Technology, Cambridge, Massachusetts 02139, USA
}

\date{\today}
\begin{abstract}
Single-shot readout is a key component for scalable quantum information processing. However, many solid-state qubits with favorable properties lack the single-shot readout capability. One solution is to use the repetitive quantum-non-demolition readout technique, where the qubit is correlated with an ancilla, which is subsequently read out. The readout fidelity is therefore limited by the back-action on the qubit from the measurement. Traditionally, a threshold method is taken, where only the total photon count is used to discriminate qubit state, discarding all the information of the back-action hidden in the time trace of repetitive readout measurement. Here we show by using machine learning (ML), one obtains higher readout fidelity by taking advantage of the time trace data. ML is able to identify when back-action happened, and correctly read out the original state. Since the information is already recorded (but usually discarded), this improvement in fidelity does not consume additional experimental time, and could be directly applied to preparation-by-measurement and quantum metrology applications involving repetitive readout.
\end{abstract}
\maketitle
\end{CJK*}

\section{Introduction}\label{sec:intro}
\begin{figure*}[t]
\begin{center}
\includegraphics[width=0.8\textwidth]{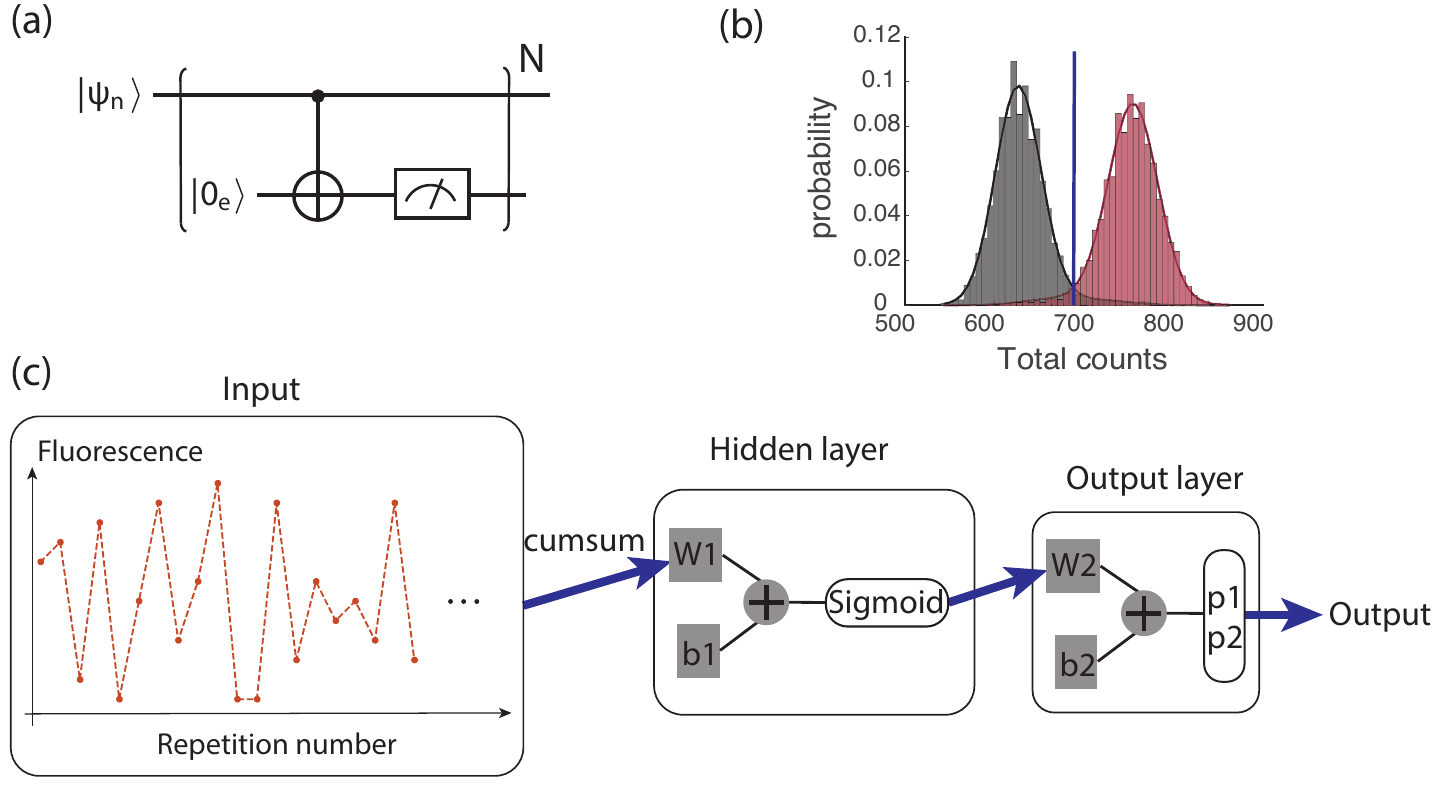}
\caption{(a) Quantum circuit for repetitive quantum-non-demolition readout of the nuclear spin state $\ket{\psi_n}$, using the ancilla electronic spin ($\ket{0_e}$). 
(b) A typical histogram of total photon numbers collected from repetitive readout, originating from bright (red) and dark (grey) states is generated using simulation and shown. A threshold at the cross point classifies future readout results in the threshold method. 
(c) Shallow neuron network architecture of MATLAB\textsuperscript{\textregistered} Neural Net Pattern Recognition tool (\textsf{nprtool}), with sigmoid as activation function and softmax output.  \textsf{nprtool} only allows users to change the number of  neurons in the hidden layer. 
The ML input is the time trace of single photon detector clicks in individual repetitive readout experiment, and we take the cumulative sum (``cumsum'') {of individual time traces} before feeding the data to the neural network. W1 (W2) and b1 (b2) are the weights and bias of the hidden (output) layer, which are learnable parameters of the network. The output is the probability p1 (p2) of the state being dark (bright).
}
\label{fig: introduction}
\end{center}
\end{figure*}

Single-shot readout is a key component for scalable quantum information processing~\cite{Divincenzo00,Raussendorf03}, for its close connection to state initialization and fault-tolerant quantum error correction~\cite{Nielsen00b}. Indeed, it is one of the main deciding factors in the selection of potential qubits. 
Single-shot readout has been achieved in various physical qubit systems, ranging from neutral atoms~\cite{Bakr09,Endres16,Cooper18}, to trapped ions~\cite{Myerson08}, superconducting qubit~\cite{Jeffrey14}, and solid-state defect centers~\cite{Morello10,Elzerman04,Hanson05,Neumann10b,Maurer12,Dreau13,Waldherr14,Liu17}.  
There are however situations where a candidate qubit has favorable coherence properties, but does not naturally come with single-shot readout capabilities. Examples include Al$^+$ ions~\cite{Schmidt05,Hume07} 
and room-temperature nitrogen-vacancy (NV) centers in diamond~\cite{Neumann10b,Maurer12,Dreau13,Waldherr14,Liu17}, where a closed optical cycle for readout is either lacking, or experimentally challenging. 
A solution to this problem is through repetitive quantum-non-demolition (QND) measurements~\cite{Hume07}.

In the repetitive QND protocol, a Controlled-NOT (CNOT) gate is applied to correlate the qubit state to an ancilla, which is subsequently read out (Fig.~\ref{fig: introduction} (a)). If the readout operator commutes with the qubit's intrinsic Hamiltonian, in other words, if the readout is QND, one can repeat the above process multiple times to increase signal-to-noise ratio, until the desired fidelity is reached. 

This protocol is also known as the repetitive readout technique widely adopted in NV research at room-temperature, where the nuclear spin state (here the \Nit or a \carb) is repetitively read out with the help of the NV electronic spin~\cite{Jiang09,Neumann10b}.
In its implementations so far, the spin state was determined by comparing the total photon number collected through all the repetitive readouts with a previously established \textit{threshold} (Fig.~\ref{fig: introduction} (b)). The detected photon count numbers are thus divided into two classes,  referred to as bright (dark) state of the qubit.

In this threshold method (TM), the readout infidelity can be evaluated from the overlap  between the photon count distributions of bright and dark states. Two factors contribute to this overlap: inefficient optical readout, including photon shot noise and limited photon collection efficiency; and deviation from the QND condition. 
The first factor can be improved by embedding the emitter into photonic structures and by using better single photon detectors. 
The second factor imposes a more fundamental constraint.
Indeed, if the readout operator does not fully commute with the system Hamiltonian, back-action from the measurement will eventually limit the number of photons that can be collected before quantum information is destroyed~\cite{Cujia19,Pfender19}. 

To mitigate this effect, we propose to use the additional information carried by the measurement-induced state perturbation itself. Information about the perturbation is already recorded during typical experiments, in the form of the time trace of photon clicks from the repetitive readouts (Fig.~\ref{fig: introduction} (c)), but is usually discarded in the TM after extracting the total photon number. Identifying the perturbation and tracing back to the unperturbed original state using this information is the key to improving the fidelity of readout.

Unfortunately, finding an elegant analytical approach proves difficult--the complexity of the photodynamics exhibits intrinsic randomness, and the inefficient photon collection process yields noisy data, precluding clean analytical analysis that would take advantage of the additional information. 
On the other hand, machine learning (ML) is designed to discover hidden data correlations, and it is widely used in classification problems~\cite{Krizhevsky17}. It has been recently introduced in quantum information tasks to mitigate crosstalks in multi-qubit readout~\cite{Seif18}, to enhance quantum metrology~\cite{Santagati19,Dinani19}, and to identify quantum phases~\cite{Lian19}.

In this work, we apply ML to  state discrimination  for the repetitive readout of NV center. To design and evaluate the ML method, we use the full information from  time trace data generated by quantum Monte-Carlo simulation. We tried different supervised ML methods and mainly focused on a shallow neural network realized using MATLAB\textsuperscript{\textregistered} Neural Net Pattern Recognition tool (npartool).
We observed consistent increase in readout fidelity using ML over TM. The improvement in readout fidelity albeit small is robust over a parameter space that covers individual NV differences. One application of our results is in preparation-by-measurement: when one discards less trustworthy measurements, ML yields a more efficient initialization process than TM.

Since in our method the training labels are readily available in experiments with very high fidelity~\cite{Neumann10b,Maurer12,Waldherr14,Dreau13,Liu17}, it can be readily applied to current experiments. Together with the robustness of our method over NV photodynamic parameters, we expect that the improved readout fidelity can be achieved in experiments.

\begin{figure*}[t]
\centering
\includegraphics[width=0.8\textwidth]{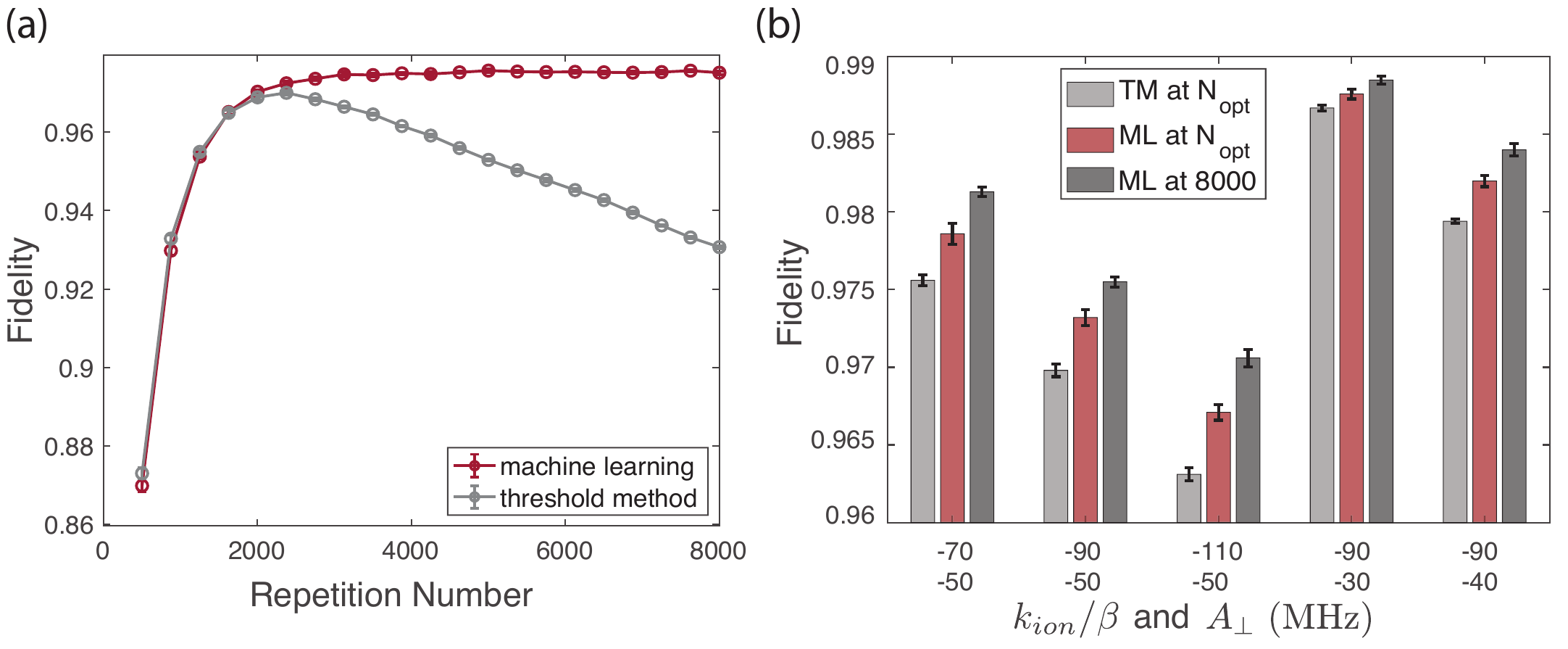}
\caption{(a) Readout fidelity as a function of repetition number $N$ in the repetitive readout. The fidelity from TM (grey) declines after $N_\mathrm{opt}=2375$ due to increasing probability of \Nit nuclear spin flips. The fidelity from ML keeps improving, although the increase rate slows down. For each repetition number, we retrain the network and take the average fidelity over 10 trainings. Error bars are the standard deviation of the 10 training results and are smaller than markers. Simulation parameters: \{$k_\mathrm{ion}=-90 \beta$MHz, $A_{\perp}=-50$MHz\}. (b) Fidelity comparison of TM at its optimal repetition number $N_\mathrm{opt}$, ML at $N_\mathrm{opt}$, and ML at $N=8000$ under different NV parameters.  $N_\mathrm{opt}$ for each were respectively (from left to right): 2000, 2375, 2750, 3125 and 2750. Error bars are the standard deviation of 10 training results. }\label{fig:mainresults}
\end{figure*}
\section{Repetitive Readout Model and Simulation}

We consider reading out the native \Nit nuclear spin state through the electronic spin of NV center at room-temperature as an example. 
The NV center's ground state is an electronic spin triplet ($S=1$), and can be optically polarized to the $\ket{m_s=0}$ state. 
The other two sublevels $\ket{m_s=\pm 1}$ have additional non-radiative decay channels under optical illumination, allowing optical readout of spin state by fluorescence intensity. 
The native \Nit nuclear spin is a nuclear spin-1 ($I=1$), and couples to the NV center through hyperfine interaction. 
\Nit does not have optical readout, but it supports a C$_\textrm{n}$NOT$_\textrm{e}$ operation (control on nuclear spin and NOT gate on electronic spin): $\ket{m_s=0,m_I=+1}\leftrightarrow\ket{m_s=+1,m_I=+1}$, and $\ket{m_s, m_I=0,-1}\leftrightarrow \ket{m_s, m_I=0,-1}$, which correlates the \Nit  to the NV state.

In the repetitive readout protocol, the NV starts in $\ket{m_s=0}$, and a CNOT gate correlates the nuclear spin state to NV. A green laser then reads out the NV state, while also  repolarizing it back to $\ket{m_s=0}$. 
Under high magnetic field, where the NV and \Nit energies are well separated, this process is approximately QND and can be repeated a few thousand times to accumulate signal, discriminating the bright $\ket{m_I=0,-1}$ (dark $\ket{m_I=+1}$) state of \Nit in a single shot~(Fig.~\ref{fig: introduction}). Still, the high magnetic field cannot fully eliminate back-action of the measurement on \Nit, which is caused by the relatively strong excited state transverse hyperfine interaction $A_\perp(S_+I_-+S_-I_+)$. 
This perturbation causes flip-flips between NV and the \Nit destroying the quantum information lost. In the TM, this perturbation prevents us from keeping to accumulate useful signal and reduces the fidelity of state discrimination.  ML, instead, as we find out, can identify the majority of such flips and therefore improve the readout fidelity. Ultimately, the readout fidelity is limited by flips that occur very early during repetitive readout.

We used simulated data to explore the effectiveness of ML in repetitive readout and to better analyze the source of improvement.
To fully capture the photodynamics involved in the repetitive readout process, we employed a 33-level model, considering the \NVm~electronic and \Nit~nuclear spins and the neutrally charged \NVz~ state. The model is described in more detail in the Appendix. Most transition rates in the model were accurately measured from independent experiments~\cite{Robledo11b,Tetienne12,Gupta16,Manson06} and we use values from Gupta et al~\cite{Gupta16}. 
The excited state NV-\Nit~transverse hyperfine interaction strength and NV$^-$ to \NVz~(de)ionization rate at strong laser power were not precisely determined before, and therefore a reasonable range is explored to cover possible variations in individual NVs, based on the results from~\cite{Maurer12,Neumann10b,Poggiali17,PhysRevB.79.235210}.

In the simulation, we assumed an intermediate magnetic field of $7500$~G typical for repetitive readout experiments, and a photon collection efficiency of $30\%$, standard with photonic structures like solid immersion lens or parabolic mirrors on the diamond~\cite{Marseglia11,Robledo11,Wan18}. A perfect CNOT gate connecting $\ket{m_s=0,m_I=+1}\leftrightarrow\ket{m_s=+1,m_I=+1}$ was assumed. Correspondingly, the dark state is $\ket{m_I=+1}$, and bright state is $\ket{m_I=0,-1}$.

We remark that it is possible to use the same protocol to read out \carb rather than \Nit~\cite{Dreau13,Waldherr14,Maurer12,Liu17}, given well-characterized hyperfine interaction strengths~\cite{Shim13x,Rao16,Smeltzer11,Dreau12}.

\section{Neural Network Architecture}

The network in \textsf{nprtool} is a two-layer feed-forward neuron network (Fig.~\ref{fig: introduction} (c)). In all trainings, we used a training set of size $10,000$~with a random portion of $15\%$ for validation. The input data is the time trace of single photon detector clicks through the repetitive readout process (Fig.~\ref{fig: introduction} (c)). Because the total photon count is a good metric for state discrimination, we take the cumulative sum of the time trace before feeding it to the neural network.
Out of the $10,000$~data, half are dark state $\ket{m_I=+1}$, while the other half are bright with a $1:1$ ratio between $\ket{m_I=0}$~and $\ket{m_I=-1}$. After training, we used a test set of size $4,000$, which was generated in the same way as the training set but not used in training, to independently test the network. 
This process is typically repeated $10$ times and the average accuracy was used throughout this work. {Error bars represent the standard deviation of the $10$ results.}

We found that approximately $12.5$ neurons per 1000 repetitions was a good balance between the increase in fidelity and avoidance of overfitting.

\begin{figure*}[htbp]
\centering
\includegraphics[width=0.8\textwidth]{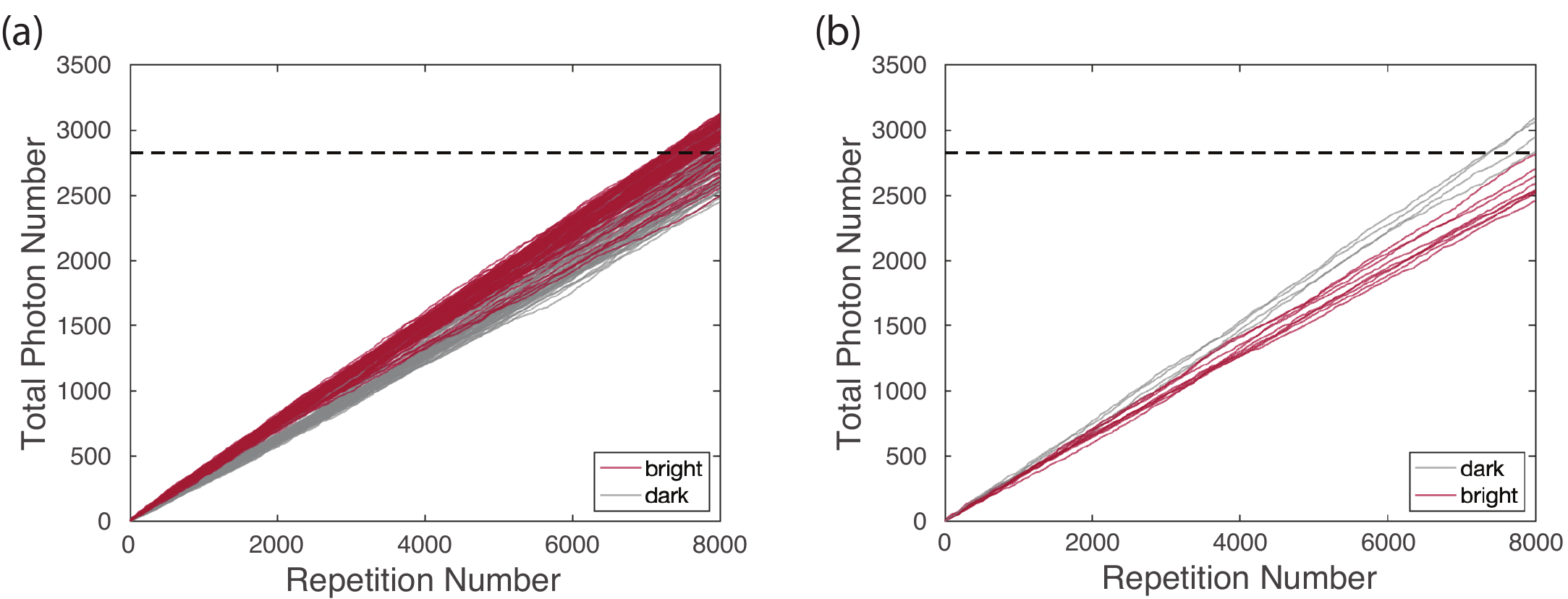}
\caption{Cumulative number of photons as a function of read out repetitions. Each trace corresponds to one input to the neural network. All traces shown here experienced at least one \Nit~flip, and are (a) correctly or (b) wrongly assigned by ML. The larger number of traces in (a) ($93.78\%$ of the total number of traces considered) reflects the high fidelity of the ML readout.
In contrast, the TM only looks at the final photon number and compares it to the threshold (dashed line), assigning roughly 25\% in (a) and all in (b) to the wrong state. 
In the figures, red lines represent time traces starting in bright state, grey in dark state; the dashed line is the threshold for $N=8000$.}\label{fig:flipflops}
\end{figure*}

\section{Results}

We first investigate the influence of the repetition number on readout fidelity. The fidelity $F$ across this manuscript is defined as
\begin{equation} \label{eqn:fidelity}
F=\frac{F_{\mathrm{bright}}+F_{\mathrm{dark}}}{2}
\end{equation}
where $F_{\mathrm{bright}}$ and $F_{\mathrm{dark}}$ are the percentage of bright and dark states that are correctly read out, respectively.

The number of repetition influences the readout fidelity in two ways: 1. A larger repetition number means more photons detected and better separation between photon count distributions of the bright (dark) states (Fig.~\ref{fig: introduction} (b)).  2. A larger repetition number, however, also implies a longer illumination time and a higher probability of the \Nit~nuclear spin to flip, due to the large transverse hyperfine interaction in the excited state, which mixes the photon count distributions of two initially different states. As a result of these competing effects, there is an optimal repetition number $N_\mathrm{opt}$ for the TM. 
On the other hand, the readout fidelity from ML keeps improving as we increase the repetition number even if the increase rate slows down (Fig.~\ref{fig:mainresults} (a)). 
At $N_\mathrm{opt}$, we observed a $0.34\%$ increase in fidelity with ML. 
Since the time trace input for ML is recorded in all experiments even when intended for TM, this improved fidelity does not consume additional experimental time. One can add more repetitions in the experiment, and harness a further increase as much as $0.57\%$ in readout fidelity (compared to TM at $N_\mathrm{opt}$). The improvement at $N>N_\mathrm{opt}$ suggests that ML is not only more robust against \Nit~flips, but rather extracts useful information from the flips. This is investigated in more detail later.

As mentioned earlier,  the excited state transverse hyperfine interaction strength $A_\perp$ between NV and \Nit~, and (de)ionization rate $k_\mathrm{ion}$($k_\textrm{deion}$) between NV$^-$ and \NVz~under strong illumination have been not yet determined to satisfactory precision. 
We therefore explored a parameter range to cover realistic values one might encounter in experiment: $A_\perp=\{-30,-40,-50\}$~MHz and $k_\mathrm{ion}=\{70,90,100\}\times\beta$~MHz, where $\beta$ is a unit-less value proportional to laser power. In the simulation, we choose $\beta$
 such that for any combination of parameters the NV would emit the same total number of photons in the bright state during repetitive readout. 
Comparisons of TM at $N_\mathrm{opt}$, ML at $N_\mathrm{opt}$ and ML at $N=8000$ are shown in Fig.~\ref{fig:mainresults} (b) under different $A_\perp, k_\mathrm{ion}$. The trend matches Fig.~\ref{fig:mainresults} (a). ML consistently outperforms TM with both repetition numbers chosen.

To better understand how ML achieves higher fidelity, we take a closer look at cases where \Nit~experienced flip-flops in the excited state, which is a major limit to the TM fidelity. 
We find the neural network is able to extract information from the time trace input to recognize if a flip has occurred, and recover the original state. Such flips could bring the photon count across the threshold, yielding misclassification when using TM. This is shown in Fig.~\ref{fig:flipflops}, where we plot the cumulative sum of the time traces in cases where flip(s) occurred. 
In Fig.~\ref{fig:flipflops} (a), ML correctly assigns all these time traces to their original states, while TM looks only at the total photon count at the end and compares it to the threshold (dashed line), making $\sim 25\%$ wrong decisions. In Fig.~\ref{fig:flipflops} (b), we show instances when ML gave the wrong classification. 
We notice that in those cases, the \Nit~flip-flops  happen at the very beginning, making the time traces indistinguishable from those of the opposite initial state with no flips. There is little hope in correctly reading out these states, posing an ultimate limit to the readout fidelity.

\begin{figure}[htbp]
\centering
\includegraphics[width=\linewidth]{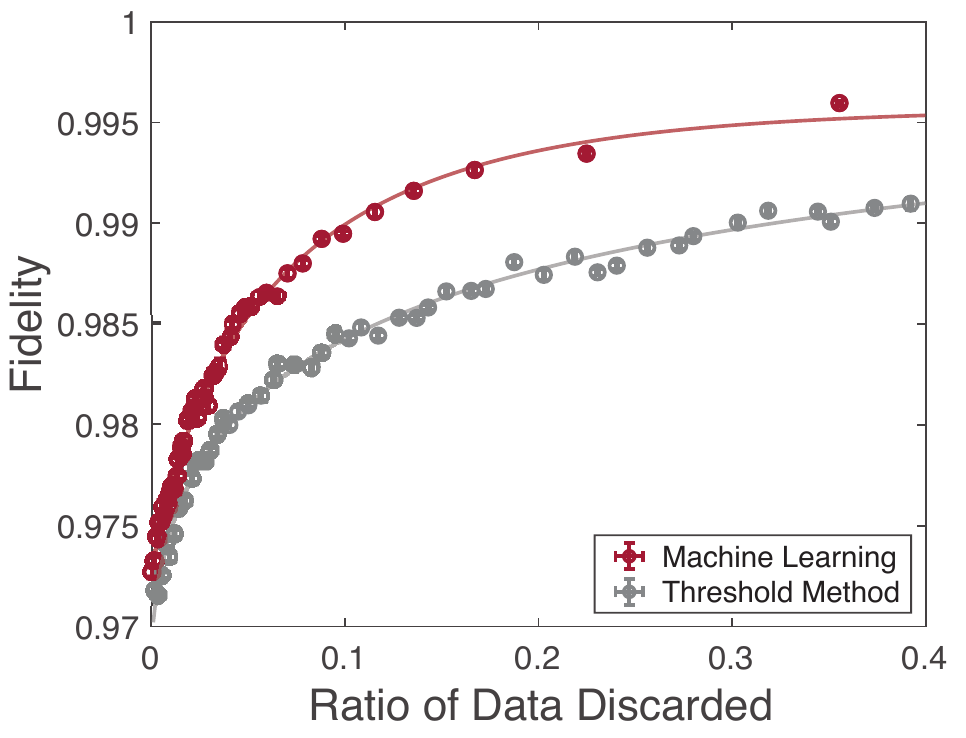}
\caption{More efficient state preparation-by-measurement. The state readout fidelity increases after discarding less trustworthy measurements and this improves the state preparation. ML always outperforms TM and scales more favorably with the ratio of discarded data. The solid curves are a guide to the eye. Error bars are the standard deviation of 10 training results, and are smaller than the marker.}\label{fig:discard}
\end{figure}

\section{Application to initialization by readout}
\label{sec:discard}
One scenario where even a modest increase in the fidelity can be beneficial is in state preparation-by-measurement~\cite{Neumann10b,Maurer12,Dreau13,Waldherr14,Liu17}. In this is a widely adopted technique, to achieve a higher fidelity of state preparation with the TM, two distinct thresholds are set, $N_\mathrm{dark}<N_\mathrm{th}$ and $N_\mathrm{bright}>N_\mathrm{th}$, where $N_\mathrm{th}$ is the readout threshold.
Measurements in between the two thresholds are discarded, as they cannot be assigned to either bright or dark state with enough confidence. This leads to a lengthier state preparation routine. 
In ML, the neural network assigns each input to a probability $p_\mathrm{bright}$ ($p_\mathrm{dark}$) of the state being bright (dark). A final step compares $p_\mathrm{bright}, p_\mathrm{dark}$ and classifies accordingly. 
To achieve a higher fidelity, we discard cases where $0.5-t<p_\mathrm{dark/bright}<0.5+t$, with an adjustable threshold $t$. 
We compare the state preparation fidelity from TM and ML, when discarding the same amount of data, and observe that ML maintains its advantage over TM, and scales more favorably than TM with the ratio of discarded measurements (Fig.~\ref{fig:discard}). This enables preparing a high fidelity initial state more efficiently. We observed similar improvement from unsupervised learning (see Appendix), agreeing with~\cite{Magesan15}.

\section{Conclusion and Outlook}

In conclusion, we have shown that ML techniques can exploit the hidden structure in the repetitive readout data of NV center at room-temperature to improve the state measurement fidelity. We used Quantum Monte-Carlo simulation based on a 33-level NV model to generate data for machine learning, and found improved single-shot readout fidelity over the traditional threshold method, that can be attributed to the ML ability to correctly classify a larger number of readout trajectory that are perturbed by the measurement process itself. 

While we used simulations, generally the training process does not depend on knowledge of the model. In fact, the only information required is the label for the state ($\ket{m_I=+1}$~or $\ket{m_I=0,-1}$), which is readily available in experiments by discarding less trustworthy data~\cite{Neumann10b,Maurer12,Dreau13,Waldherr14,Liu17}. 
One can then use this data to train a network specific to the NV of interest, and expect an increase in readout fidelity in all subsequent repetitive readout experiments, free of any additional experimental time.
Although individual NVs may have slightly different photodynamic parameters, they should be covered by the range we explored in this work, and therefore the improvement in fidelity is expected to be ubiquitous.

In addition, the off-the-shelf MATLAB\textsuperscript{\textregistered} deep learning toolbox we employed greatly reduces the complexities in the neuron network architecture, making this improvement easily reproducible and more accessible to experimentalists.

Though small, the increase in fidelity does not require any additional experimental time, and is readily compatible with experiments using repetitive readout of nuclear spins, including in quantum metrology~\cite{Aslam17,Lovchinsky16,Degen17} to improve sensitivity.

To further shed light on the  bright/dark decisions that affect the  ML readout fidelity, one could use decision tree learning instead of a neuron network. This could potentially inform optimized readout protocols, with varying illumination times, or help further improve the neuron network architecture.
More broadly, ML could be applied to more complex systems, for example to help mitigate crosstalk of fluorescence signals in a solid-state register consisting of a few nearby NV or other color centers~\cite{Seif18}.

\appendix
\section{Appendix I: NV model and Quantum Monte-Carlo Simulation}\label{appdix:model}
\begin{figure*}[htbp]
\centering
\includegraphics[width=0.9\textwidth]{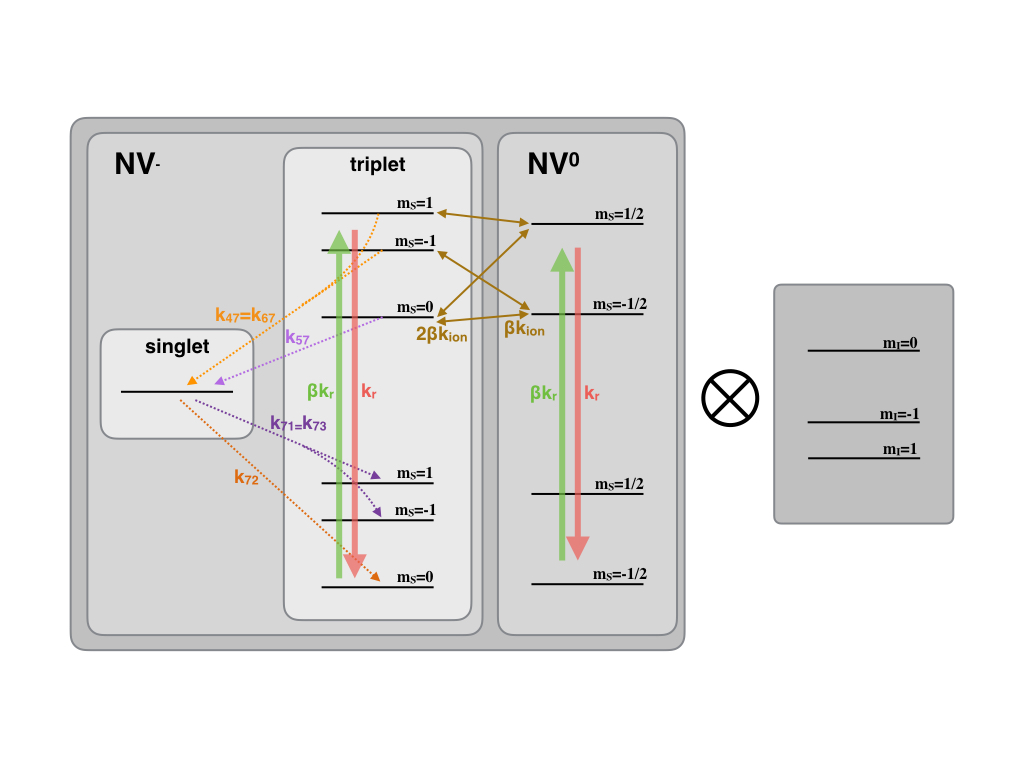}
\caption{The 33-level NV model used in our simulation, consisting of 11 electronic spin levels times 3 nuclear spin levels (level spacings not to scale). $k_r$, $k_{47}(=k_{67})$, $k_{57}$, $k_{71}(=k_{73})$, $k_{72}$ and $k_\mathrm{ion}$ are incoherent transition rates connecting the corresponding energy levels. The optical transition rate $k_r$ between excited state and ground state  are set equal for  \NVm and \NVz, and are assumed to be spin-conservative (spin non-conservative part is $<1\%$~\cite{Robledo11b}). $\beta$ is a dimensionless parameter given by the ratio of the laser power to the optical transition rate. $k_\mathrm{(de)ion}$ is the (de)ionization rate. We assume the (de)ionization happens in the excited state and follows the selection rules depicted by the brown arrows.}\label{fig:model}
\end{figure*}

We used a 33-level model to fully describe the dynamics of NV-\Nit~in the repetitive readout process. This model includes the spin-1 triplet ground and excited states, and singlet metastable state for \NVm, the spin-$1/2$ ground and excited states for \NVz, and the nuclear spin-1 of \Nit, as illustrated in Fig.~\ref{fig:model}. The transition rates directly related to the NV photoluminescence  have been precisely determined and reported in various works~\cite{Robledo11b,Tetienne12,Gupta16,Manson06}, although with some significant variations. 
For the simulation we took the values from Gupta \textit{et al.}~\cite{Gupta16} listed in Table.~\ref{tab:transitionrates}. 

\begin{table}[h]
\begin{tabular}{l|l|l|l|l|l|l}
transition rates& $k_r$ & $k_{47}$ & $k_{57}$ & $k_{71}$ & $k_{72}$ \\
\hline
(MHz)& 65.9 & 92.1 & 11.4 & 1.18 & 4.84 \\
\hline
\end{tabular}
\caption{Transition rates used in the 33-level model.}\label{tab:transitionrates}
\end{table}

The exact (de)ionization mechanisms under $532$~nm laser illumination have not been yet determined experimentally, neither have the (de)ionization rate under laser-power comparable to the saturation power (measurement under weak power can be found in ~\cite{Aslam13,Chen13b,Hacquebard18}). Here we assume the (de)ionization $k_\textrm{ion}$($k_\textrm{deion}$) occurs only in the excited states, and obeys  selection rules as illustrated in Fig.~\ref{fig:model}. To maintain the experimentally determined $70/30$ ratio~\cite{Aslam13} between the charge states, we set $k_\textrm{deion}=2k_\textrm{ion}$. The ionization rate is proportional to the  laser intensity, which is swept around $k_\textrm{ion}\approx 90\beta$~MHz, in accordance with \cite{Maurer12}.

When the magnetic field is applied along the NV-axis, the ground state NV-\Nit~Hamiltonian has negligible effect on the repetitive readout, thus it is not considered in the numerical simulation. The \NVm~excited state Hamiltonian reads:
\begin{equation}\label{eqn:NV-ham}
H_{-}=\Delta_{es} S_z^2+Q I_z^2+\gamma_e B S_z+\gamma_n B I_z+\mathbf{S}\cdot\mathbf{A}\cdot\mathbf{I}
\end{equation}
where $\mathbf{S}$ and $\mathbf{I}$ are the electronic and nuclear spin operators, $\Delta_{es}=1.42$~GHz is the zero-field splitting of the electronic spin, $Q=-4.945$~MHz the nuclear quadrupole interaction~\cite{Smeltzer09}, and $\gamma_e=2.802$~MHz/G and $\gamma_n=-0.308$~kHz/G the electronic and nuclear gyromagnetic ratios. The hyperfine interaction term is diagonal due to symmetry:
\begin{equation}\label{eqn:NV-hfgs}
\mathbf{S}\cdot\mathbf{A}\cdot\mathbf{I}=A_\parallel S_zI_z+A_\perp(S_xI_x+S_yI_y)
\end{equation}
where $A_\parallel=-40$~MHz were determined via ODMR experiment~\cite{Neumann09}. $A_\perp$ was believed to be similar to $A_\parallel$ and is recently measured between $-40$ and $-50$~MHz~\cite{Poggiali17}. 

The \NVz~excited state Hamiltonian takes the form:
\begin{equation}\label{eqn:NV0ham}
H_{0}=Q I_z^2+\gamma_e B S_z+\gamma_n B I_z+\mathbf{S}\cdot\mathbf{C}\cdot\mathbf{I}
\end{equation}
with the hyperfine interaction term:
\begin{equation}\label{eqn:NV-hf}
\mathbf{S}\cdot\mathbf{C}\cdot\mathbf{I}=C_\parallel S_zI_z+C_\perp(S_xI_x+S_yI_y)
\end{equation}
The hyperfine interaction strengths were considered similar to those in the \NVm~excited state~\cite{PhysRevB.79.235210}, and we set $C_\parallel=C_\perp=-40$MHz. 

To simulate repetitive readout experiments for both the training and testing data, we used the quantum Monte-Carlo method based on the aforementioned 33-level model. One challenge lies in the various time scales involved in the numerical simulation, from the electronic spin's fast oscillation $\omega\sim (2\pi)\cdot 10$~GHz, to the optical transition rates $k_{ij}\sim 100$~MHz, to the flip-flop rate of \Nit~nuclear spin $1/T_1^n \sim $~kHz. We mitigate this issue by employing the Born-Oppenheimer approximation~\cite{doi:10.1002/andp.19273892002} in our numerical simulation, and average out the fast oscillation at $\omega$ as following.

We define $\delta p_{mn}$ as the transition probability from the state $\ket{m}$ to $\ket{n}$ in the time step $\delta t$. 
Starting from $\ket{\psi(t=0)}=\ket{m}$, we have
\begin{equation}\label{eqn:boApprox}
\begin{aligned}
\delta p_{mn} &=\int^{\delta t}_0 \left( \sum^{33}_{i=1}|\braket{n}{i}|^2~|\braket{i}{\psi(t)} |^2 \right)dt \\ & = \sum^{33}_{i=1} \left( k_{in}\int^{\delta t}_0 |\braket{i}{\psi(t)}|^2 dt \right)
\end{aligned}
\end{equation}
Notice that $\vert\braket{i}{\psi(t)}\vert^2$ is periodic with period $2 \pi/ \omega$, which is much smaller than the time step $\delta t \sim 1/k_{ij}$. Thus, we assume only the average effect of this oscillation is seen in each time step, and numerically find $\left\langle \frac{\delta p_{mn}}{\delta t}\right\rangle$. This allows us to efficiently perform the quantum Monte-Carlo simulation.

\section{Appendix II: Machine Learning Discussions}\label{appdix:MLdiscussion}
\subsection{Recurrent Neural Network}
Recurrent neural network (RNN) is a commonly used architecture specializing in time-series data with the capability to understand the correlation within the time-series. In the main text, we showed results obtained using shallow neural network. In order to see if we gain by exploiting the correlation within the time series we also tested the performance of an advanced recurrent neural network: long short-term memory (LSTM).
 Due to the nature of recurrent neural network, the training process is very time-consuming and therefore not suitable for exploring multiple parameters in our model. To speed up the training process, we averaged the input time trace data over 100 realizations, to greatly reduce the training set dimension. Indeed, this may have caused some loss of information. The result though still consistently outperforms the TM and is comparable to the shallow neural network shown in the main text (see Table.~\ref{tab:LSTM}). One remark is that we did not take the cumulative sum for the input data, because LSTM specializes in time series data and is able to recognize some quasi-periodic patterns.

\begin{table}[htbp]
  \centering
    \begin{tabular}{|c|c|c|c|c|}
    \hline
   $A_\perp$ (MHz) & $k_\mathrm{ion}$ (MHz) & TM fidelity & ML fidelity & LSTM fidelity \\
   \hline
    \multicolumn{1}{|c|}{\multirow{3}[0]{*}{-50}} & \multicolumn{1}{c|}{$70\beta$} & $97.56(4)$\% & 97.86$(7)$\% & 97.61$(5)$\% \\
          & \multicolumn{1}{c|}{$90\beta$} & 96.98$(4)$\% & 97.32$(5)$\% & 97.40$(2)$\% \\
          & \multicolumn{1}{c|}{$110\beta$} & 96.31$(4)$\% & 96.71$(5)$\% & 96.77$(7)$\% \\
    \hline
    $-30$   & \multirow{3}[0]{*}{$90\beta$} & 98.67$(2)$\% & 98.76$(3)$\% & 98.44$(3)$\% \\
    $-40$   &       & 97.94$(2)$\% & 98.20$(4)$\% & 98.29$(3)$\% \\
    $-50$  &       & 96.98$(4)$\% & 97.32$(5)$\% & 97.40$(2)$\% \\
    \hline
    \end{tabular}%
      \caption{Comparison between the fidelity obtained through TM, ML and LSTM under different parameters. All training and testings were conducted at the $N_\mathrm{opt}$ of that set of parameters. Overall, the LSTM algorithm has similar performance compared with the shallow neural network.}
  \label{tab:LSTM}%
\end{table}%

\subsection{Unsupervised learning}
In the main text we compared the enhanced fidelities of TM and supervised learning after discarding less trustworthy data. Another possibility is to use unsupervised learning~\cite{Magesan15}. This method is of interest because unsupervised learning does not require any well-labelled data. We implemented the \textsf{k-means} algorithm that classifies a given data set into $k$ different groups.

\begin{figure}[htbp]
\centering
\includegraphics[width=0.9\linewidth]{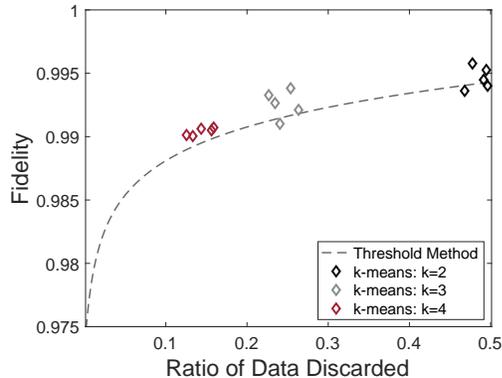}
\caption{More efficient state preparation-by-measurement.
Improved dark state readout accuracy after discarding less trustworthy readouts. Each diamond-shaped point represents an individual k-means test.}\label{fig:unsupervised_discard}
\end{figure}

We first use the TM readout  to obtain a bright (dark) group of measurement trajectories. We then perform \textsf{k-means} on the bright (dark) group to further classify it into $k$ subgroups. 
The fidelity increases when we discard the smallest subgroup. Compared to the TM, \textsf{k-means} gives better  fidelity as shown in Fig.~\ref{fig:unsupervised_discard}, because the unsupervised learning  extracts some information about \Nit flips through the hidden structures in time trace data, in agreement with~\cite{Magesan15}.   
Note that unlike TM or supervised learning, we cannot control the ratio of discarded data. Therefore, the fidelity defined by Eq.~\ref{eqn:fidelity} is not available, and only the fidelity of dark state is shown. We also remark that in rare cases, \textsf{k-means} gives outlier results with fidelity much worse than TM. 

\subsection{Robustness of trained network}

In addition to the universal improvement from ML over TM with different photodynamic parameters explored in the main text, we here explore the robustness of a network to  a change in parameters. In particular, we use the network R trained by \{$k_\mathrm{ion}=-90 \beta$MHz, $A_{\perp}=-50$MHz\} data, to classify data generated with different parameters.

First, we test the network R on different (de)ionization rate \{$k_\mathrm{ion}=-100 \beta$MHz, $A_{\perp}=-50$MHz\}, obtaining a fidelity of $97.93(5)\%$ from the network R, compared to $98.07(5)\%$ from TM. 
We attribute this deteriorated performance of ML to the change in the photodynamics. 
Under the same condition, different $k_\mathrm{ion}$ change the relative distributions of bright and dark states.  This change cannot be compensated by laser intensity, and makes the network R obsolete.

We then tested the network R robustness to different transverse hyperfine strengths, $A_\perp=-40, -30$~MHz. 
Intuitively, a small change in $A_\perp$ does not change the photoluminescence pattern, but rather modifies the \Nit~flip-flop rate a little, which could be captured by the network, given its ability to recognize the occurrence of flip-flops. Indeed, we observed better fidelity from the network R on $A_\perp=-40$~MHz data than TM, and comparable fidelity to TM on $A_\perp=-30$~MHz, where the parameter has changed by $40\%$ (Table.~\ref{tab:train_A_test_B}). Here we used $N_{opt}$ for the test data for both ML and network R. These results indicated that provide variations in the NV parameters are small, it is possible to use a fixed network R to directly read out any NV, without the need to run experiments to generate the traning data.

\begin{table}[htbp]
\begin{tabular}{c|c|c|c}
$A_\perp$ (MHz)& TM fidelity & ML fidelity &  network R fidelity\\
\hline
$-40$ & $97.94(2)\%$ & $98.20(4)\%$ & $98.24(4)\%$  \\
\hline
$-30$ & $98.67(2)\%$ & $98.76(3)\%$ & $98.66(4)\%$ \\
\hline
\end{tabular}
\caption{Robustness test of network R trained with \{$k_\mathrm{ion}=-90\beta$MHz, $A_{\perp}=-50$MHz\}. We compare the readout fidelities of test data with different $A_\perp$ from TM, ML, and network R. The result from network R is better than TM when $A_\perp$ is not changed too much.}\label{tab:train_A_test_B}
\end{table}

\bibliography{SingleShotML,Biblio} 
\bibliographystyle{apsrev4-1}

\end{document}